\documentclass[11pt,a4paper]{article}
\usepackage[english]{babel}
\usepackage{jheppub}

\title{Measurement of the neutrino velocity with the 
OPERA detector in the CNGS beam using the 2012 dedicated data}

\author[af]{T.~Adam,}
\author[u]{N.~Agafonova,}
\author[y]{A.~Aleksandrov,}
\author[t]{A.~Anokhina,}
\author[q]{S.~Aoki,}
\author[f]{A.~Ariga,}
\author[f]{T.~Ariga,}
\author[ai]{D.~Autiero,}
\author[ak]{A.~Badertscher,}
\author[f]{A.~Ben Dhahbi,}

% ...
\affiliation[	f	]{	Albert Einstein Center for Fundamental Physics, Laboratory for High Energy Physics (LHEP), University of Bern, CH-3012 Bern, Switzerland	}
% ...
\affiliation[	q	]{	Kobe University, J-657-8501 Kobe, Japan	}
% ...
\affiliation[	t	]{	INR-Institute for Nuclear Research of the Russian Academy of Sciences, RUS-117312 Moscow, Russia	}
\affiliation[	u	]{	ITEP-Institute for Theoretical and Experimental Physics RUS-117259 Moscow, Russia	}
% ...
\affiliation[	w	]{	(MSU SINP) Lomonosov Moscow State University Skobeltsyn Institute of Nuclear Physics, RUS-119992 Moscow, Russia	}
% ...
\affiliation[	y	]{	INFN Sezione di Napoli, I-80125 Napoli, Italy	}
% ...
\affiliation[	af	]{	IPHC, Universit\'e de Strasbourg, CNRS/IN2P3, F-67037 Strasbourg, France	}
% ...
\affiliation[	ai	]{	IPNL, Universit\'e Claude Bernard Lyon I, CNRS/IN2P3, F-69622 Villeurbanne, France	}
% ...
\affiliation[	ak	]{	ETH Zurich, Institute for Particle Physics, CH-8093 Zurich, Switzerland	}

% ...

\abstract
{
In spring 2012 CERN provided two weeks of a short bunch proton beam dedicated to the neutrino velocity measurement over a distance of 730~km.
The OPERA neutrino experiment at the underground Gran Sasso Laboratory used
an upgraded setup compared to the 2011 measurements, improving the measurement time accuracy.
An independent timing system based on the Resistive Plate Chambers was exploited providing
a time accuracy of $\sim$1 ns. Neutrino and anti-neutrino contributions were separated using the information
provided by the OPERA magnetic spectrometers.
The new analysis profited from the precision geodesy measurements of the neutrino baseline
and of the CNGS/LNGS clock synchronization. The neutrino arrival time with respect to the one computed assuming
the speed of light in vacuum is found to be $\delta t_\nu \equiv TOF_c - TOF_\nu= ( 0.6 \pm 0.4\ (stat.) \pm 3.0\ (syst.) )$~ns
and $\delta t_{\bar{\nu}} \equiv TOF_c - TOF_{\bar{\nu}} = ( 1.7 \pm 1.4\ (stat.) \pm 3.1\ (syst.) )$~ns
for $\nu_{\mu}$ and $\bar{\nu}_{\mu}$, respectively. This corresponds to a limit on the muon neutrino velocity
with respect to the speed of light of $-1.8 \times 10^{-6} < (v_{\nu}-c)/c < 2.3 \times 10^{-6}$ at 90\%~C.L.
This new measurement confirms with higher accuracy the revised OPERA result.}

\arxivnumber{1234.5678}

\begin{document}

\maketitle

\setcounter{page}{1}

\section{Introduction}
\label{int}

The OPERA neutrino experiment \cite{ref1} reported in 2011 the observation of an anomalous value of the CNGS muon neutrino time-of-flight
which could be interpreted in terms of superluminal propagation~\cite{v2}.
Continuing carefully to scrutinize the measurement, two unaccounted systematic effects significantly affecting the result were found by the Collaboration.
A final result taking into account those effects was reported in \cite{v3} where no significant difference between the speed of light and the neutrino velocity was observed.

In order to cross--check the revised result and improve the precision of the measurement, CERN provided a narrow bunch proton beam between the 10th and 24th of May 2012.
OPERA profited of this new dedicated run to improve its timing system and to further demonstrate the validity of the procedure used to correct the previously reported measurement~\cite{v3}.
In this new run, both the Target Tracker (TT) and Resistive Plate Chambers (RPC) were used independently.
These detectors are briefly described in Section~\ref{ope}.
We report here the results obtained by each sub--detector separately and their combined result.
Profiting from the information given by the muon spectrometers, results are provided separately for muon neutrinos and anti-neutrinos present in the CNGS beam.
Indeed, $\sim$4\% of the Charged Current (CC) interactions in the surrounding rock and inside the detector result from $\bar{\nu_\mu}$ interactions~\cite{antin}.

\section{The new CNGS narrow--bunch neutrino beam}
\label{cngs}

In autumn 2011 a first CNGS narrow--bunch wide--spacing neutrino beam was used in order to cross--check the neutrino velocity result previously obtained by OPERA with a statistical method applied to the data accumulated with the standard CNGS neutrino beam tuned for the $\nu_\tau$ appearance search in $\nu_\mu \rightarrow \nu_\tau$ oscillation~\cite{v3}.
This test proved that it was possible, within a short period of two to three weeks of bunched beam, to obtain similar or even better accuracy for  the neutrino velocity measurement compared to the statistical analysis applied to the data accumulated with the standard CNGS beam over years.

\begin{figure}[hbt]
\begin{minipage}{.45\linewidth}
\begin{center}
\mbox{\epsfig{file=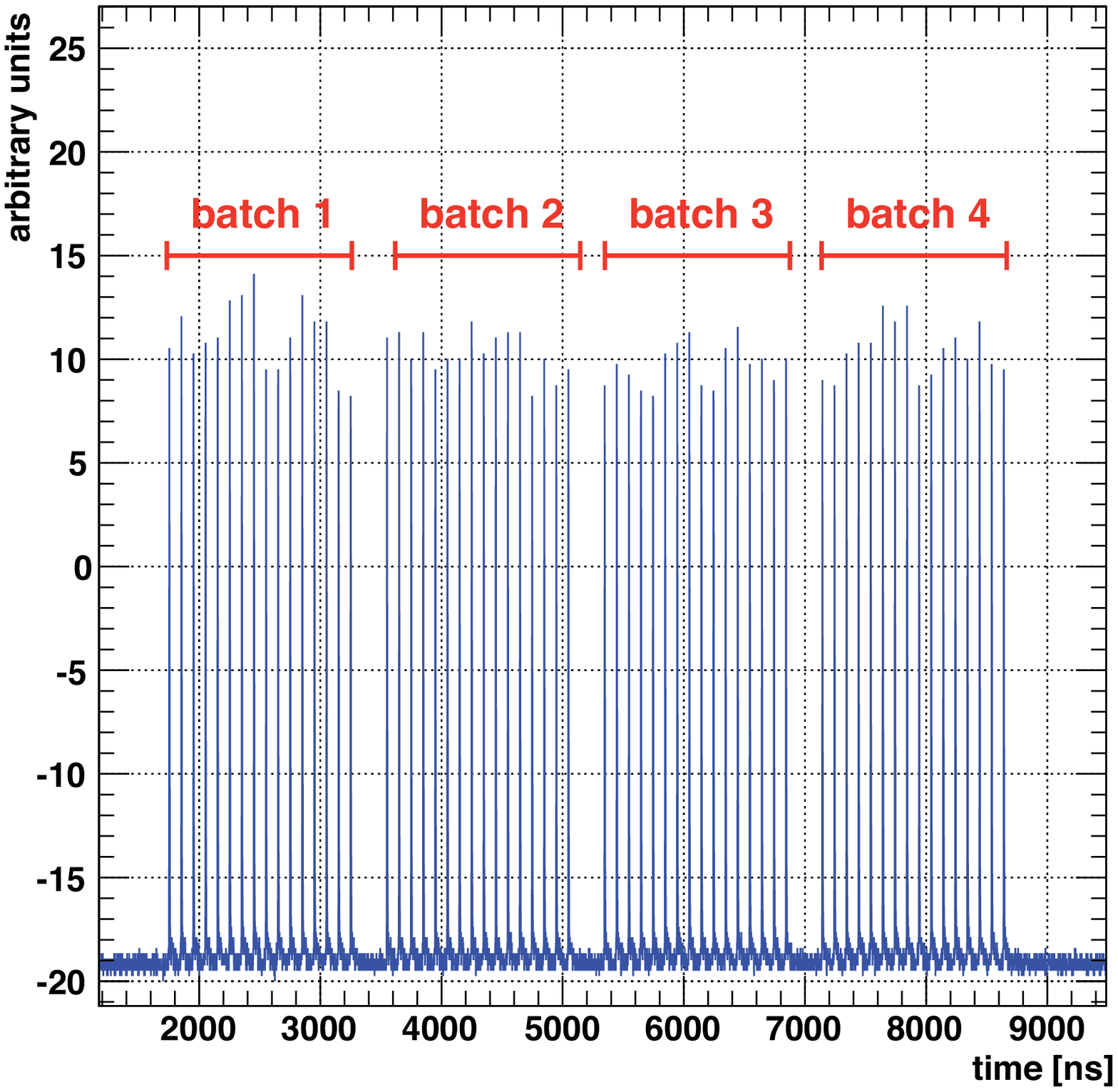,width=7cm}}
\caption{\small Intensity of the 2012 proton beam as a function of time for one SPS extraction as recorded by the BCT.} \label{bb1}
\end{center}
\end{minipage} \hspace{1.cm}
\begin{minipage}{.45\linewidth}
\begin{center}
\mbox{\epsfig{file=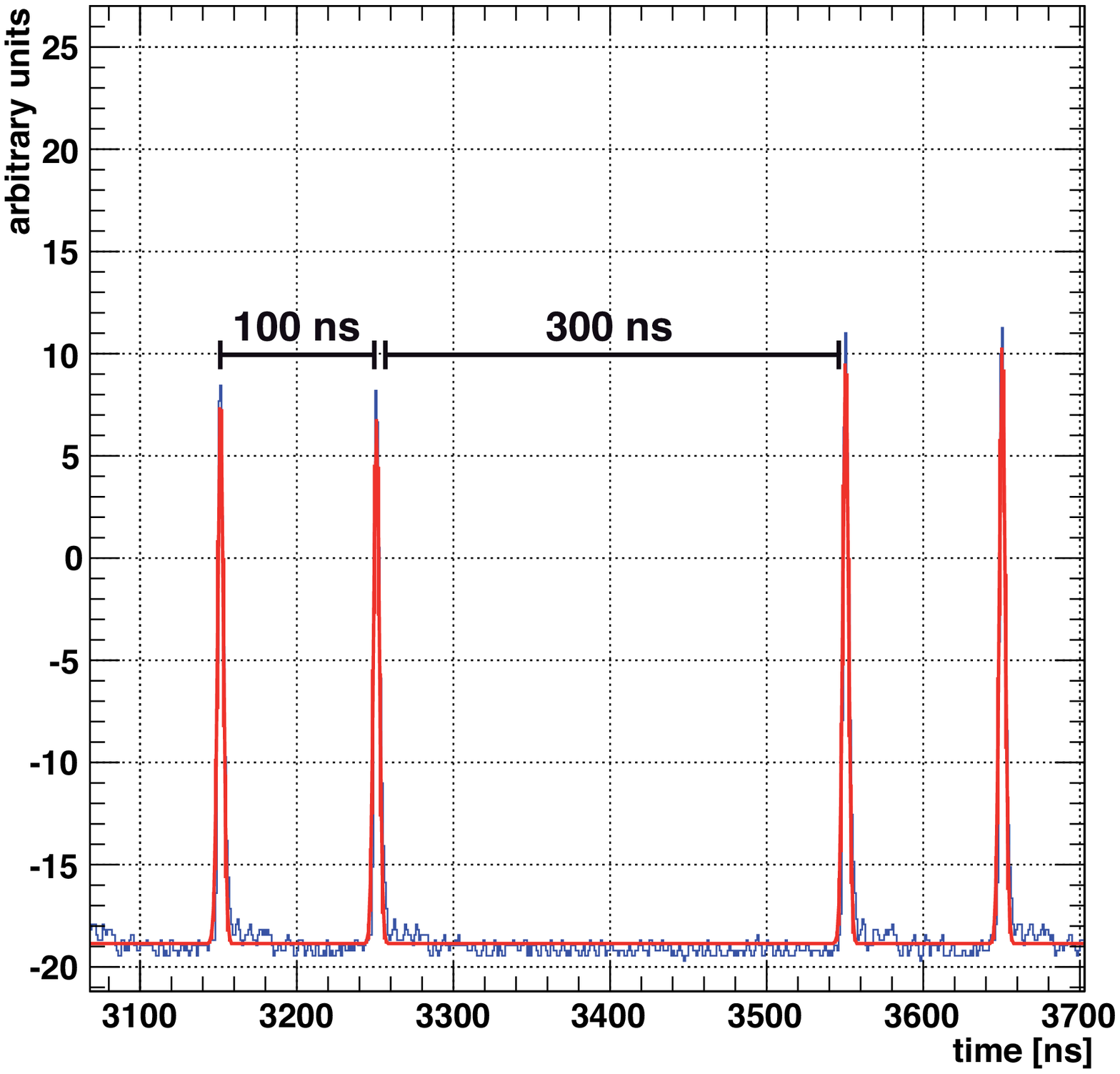,width=7cm}}
\caption{\small Zoom of Fig.~\ref{bb1} in the region between two proton batches.} \label{bb2}
\end{center}
\end{minipage}
\end{figure}

The 2011 narrow--bunch beam had bunches separated by 524~ns yielding a total intensity of $1.1\times 10^{12}$ protons per cycle.
In the 2012 run with narrow--bunch narrow--spacing beam (BB) the bunch separation is only 100~ns in order to increase the number of delivered neutrinos per time unit.
In each CNGS cycle, lasting 13.2~s, a single extraction delivers 4 batches of 16 proton bunches.
Each 1.8~ns long (RMS) bunch contains $\sim$10$^{11}$ protons providing an intensity 6 times higher than during the 2011 BB run.
Fig.~\ref{bb1} shows the proton waveform of one SPS extraction and Fig.~\ref{bb2} gives a closer view of 4 proton bunches, two at the end of the first batch and two at the beginning of the second one, as measured by a Beam Current Transformer (BCT) at CERN.
The bunch spacing of 100~ns still allows to uniquely identify the proton bunch corresponding to a given neutrino interaction in the OPERA detector.

In two weeks run a total of 1.8$\times$10$^{17}$ protons on target (p.o.t.) was delivered. OPERA recorded 104 on--time events, 67 involving the TT detectors and 62 involving the RPC detectors.

Just before this running period CERN installed a monitoring system, called White Rabbit~\cite{wr}, at CERN and at LNGS to monitor relevant timing parameters of common systems and also specific parameters of each LNGS experiment.
OPERA only used this facility to monitor the delay of the external GPS time signal along 8.3~km of the LNGS optical fibre and the frequency of the Master Clock (MC) oscillator which compromised the measurements performed in 2011.

\section{The OPERA detector and its timing systems}
\label{ope}

The OPERA neutrino detector at LNGS is composed of two identical Super Modules, each consisting of an instrumented
target section with a mass of about 625 tons followed by a magnetic muon spectrometer.
Each target section is a sequence of walls filled with emulsion film/lead plate modules interleaved with pairs of horizontal and vertical $6.7 \times 6.7$~m$^2$ planes of 256  scintillator strips composing the TT~\cite{ref9}.
The TT allows the location of neutrino interactions in the target.
This detector is also used to measure the arrival time of neutrinos.
The scintillating strips are read out on both sides through wavelength shifting fibres coupled to 64-channel photomultipliers.
Each of the two OPERA magnets is instrumented with 22 planes of RPC's~\cite{ref1}. 
Each plane can provide the $x$ and $y$ transverse coordinates for crossing particles using horizontal and vertical readout strips.

The DAQ records the timestamp of the earliest TT photomultiplier signal of the event and the earliest hit of each RPC module reaching the readout electronics.
A time calibration of these two sub-detectors allowed converting the measured time in UTC.
The complete OPERA timing system is extensively described in~\cite{v3}.
Here, only improvements made for the new BB run are presented (faster optical/electrical converter, internal Master Clock TDC, RPC Trigger Board timing).

A faster optical/electrical converter was used in order to avoid the effect of time--walk discussed in~\cite{v3} caused by the low light intensity of the 1PPmS GPS signal after 8.3~km path along the optical fibre from the external to the underground laboratory.
After the optical/electrical converter processing, the 1PPmS signal has a jitter of 3.2~ns (RMS) w.r.t. the input signal.
The previous measurements of the neutrino velocity were affected by a jitter of $\pm 25$~ns due to the 1PPmS signal tagging with respect to the uncorrelated internal frequency of the 20~MHz MC.
As a consequence, the timestamp of all hits recorded in the same 0.6~s DAQ cycle (the reset of the cycles is sent to all OPERA sensors every 600 1PPmS pulses) was shifted by the same quantity.
In order to remove this jitter an internal TDC of the MC FPGA with an accuracy of $\sim$300~ps was used to tag the arrival of the external 1PPmS with respect to the start of the internal oscillator time--bins, 50~ns wide.

\begin{figure}
  \centering
 \label{cycles}
  \includegraphics[height=.25\textheight]{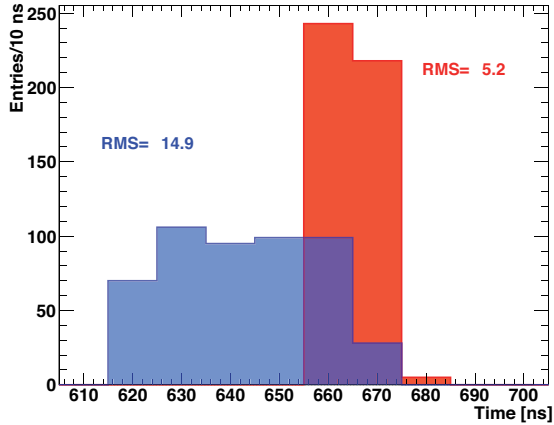}
  \caption{Recorded time for 1PPmS pulses injected at the level of one TT channel and coinciding with the DAQ reset, before (blue) and after (red) TDC correction.}
\end{figure}

Because of the 3.2~ns jitter of the 1PPmS signal, hits occurring near the end of the 50~ns time--bins of the internal oscillator are sometimes incorrectly associated with the next time--bin.
The corresponding event timestamps are thus shifted by 50~ns.
The fraction of shifted events (outliers) is 5.2\%.
These outliers are removed by rejecting cases where the TDC time (varying from 0 to 50~ns) exceeds 45~ns.
Fig.~\ref{cycles} shows the time distribution with respect to the beginning of each DAQ cycle for 1PPmS pulses coinciding with the DAQ reset sent to one TT channel through an injection capacitor.
The timing distribution as recorded by the DAQ without TDC corrections is shown in blue, the same distribution after correction is given in red.
 The improvement on the time resolution when using TDC corrections is clearly seen.
 The same distribution was also used to accurately estimate the time delay between the MC and the reference TT sensor (the time shift between the two distributions is arbitrary).

The RPC timing system is described in \cite{v3}.
For these new measurements, in order to reduce the systematic errors, an 
accurate determination of all relevant delays was done.
The starting points of the neutrino interaction time reconstruction 
are the RPC hits associated with the reconstructed muon tracks.
The maximum number of hits for tracks crossing one complete spectrometer is 22
(44 hits for tracks crossing both spectrometers).
The time of the signal formation in the RPC gas was obtained from measurements performed on a detector 
prototype.
The delays related to the signal propagation in the readout strips up to 
the front end boards and the signal processing in the readout electronics up to the timestamp 
digitization delays were obtained from a dedicated in situ time calibration campaign~\cite{rpc}.
The 1PPmS pulse of the LNGS atomic clock was sent through the same readout chain as the one used for the neutrino data.
Values are summarized in Tab.~\ref{sys_tab}.

\begin{table}
  \centering
\caption{Parameters used and related systematic errors on the measurement of $\delta t$.}
\begin{scriptsize}
\begin{tabular}{llr}
\hline
 \bf{Sub-system} &\bf{Parameter} & \bf{ns}  \\
\hline
Common                  & Baseline ($\pm 20$~cm)   & 731278~m $\pm 0.7$    \\
Common                  & FWD trigger delay              & $\pm 1.0$                          \\
Common                  & BCT calibration                   & $583.7\pm 1.0$                \\
Common                  & CNGS-OPERA GPS synchronisation   & $2.3\pm 1.7$  \\
Common                  & UTC delay                            & $\pm 1.0$                              \\
\hline
Common        & \bf{Total common systematic uncertainty}   & \bf{$\pm 2.5$}  \\
\hline
\hline
TT+RPC (MC)       & LNGS fibre                           & $41067\pm 1.0$               \\
For RPC-TB             & LNGS fibre                           & $41319.5\pm 0.2$               \\
TT+RPC (MC)       & DAQ clock transmission    & $7046\pm 1.0$                 \\
\hline
\hline
Common TT        & FPGA calibration                & $24.5\pm 1.0$                   \\
Common TT        & Time response                   & $49.3\pm 0.9$                   \\
\hline
TT Method 1        & MC simulation                    & $9.4\pm 3.0$                      \\
\hline
TT Method 1        & \bf{TT total intrinsic 1st-hit systematic uncertainty}   & \bf{$\pm 3.3$}   \\
\hline
TT Method 2        & MC simulation                    & $\pm 0.7$                      \\
\hline
TT Method 2        & \bf{TT Total intrinsic $\mu$ systematic uncertainty}   & \bf{$\pm 1.5$}   \\
\hline
\hline
Common RPC      & Signal formation    & $24.0 \pm 2.0$                 \\
Common RPC      & Strip propagation delay (4.54 ns/m)    & $ \pm 0.1$                 \\
Common RPC      & Interconnection board                & $2.4\pm 0.1$                   \\
\hline
RPC Method 3        & FEB delay                & $\pm 1.0$                   \\
RPC Method 3        & Internal mezzanine delay         & $\pm 1.0$                      \\
RPC Method 3        & Non-uniformities         & $\pm 3.9$                   \\
\hline
RPC ($\mu$)       & \bf{RPC Total intrinsic  $\mu$ systematic uncertainty}   & \bf{$\pm 4.6$}   \\
\hline
RPC Method 4        & electronic chain delays                    & $228.0\pm 0.8$                      \\
RPC Method 4        & TDC ref. frequency               & $\pm 0.1$                   \\
RPC Method 4        & TDC meas. ref. time               & $\pm 0.7$                  \\
RPC Method 4        & TDC integral non-linearity    & $\pm 0.3$                  \\
\hline
RPC Method 4       & \bf{RPC Total intrinsic TB systematic uncertainty}   & \bf{$\pm 2.3$}   \\
\hline
\end{tabular}
\end{scriptsize}
\label{sys_tab}
\end{table}

\label{rpc_tb}

The OPERA spectrometers are equipped with a new dedicated timing system using the purposely developed electronics for triggering the Precision Tracker (PT) drift tubes: seven RPC layers in each magnet are instrumented with 
timing boards~\cite{tb_pnote}, each one discriminating the 
positive polarity signals and forming the OR from 16 read-out strips.
Each RPC layer is served by 14 timing boards (TB), whose digital output
signals are sent through 16~m long flat cables to the OR Plane Electronics (OPE) boards.
Each OPE board forms the OR signal of an entire RPC layer which is further
used to trigger the PT~ \cite{trigger_pnote}.
An additional single--ended positive polarity output is present in the OPE
and it was used for the new timing system.

The UTC time of neutrino interactions with muons crossing the OPERA
spectrometers is reconstructed in two steps: 
the UTC time coded in the synchronization signal (1PPmS), sent by the ESAT 
LNGS atomic clock, is acquired, while the sub-ms component is
computed as a time difference between the event hits registered by the 
TB and the leading edge of the 1PPmS measured by means of a TDC.

The decoding of the 1PPmS UTC time is performed by a custom VME module,
called slave clock, which also reproduces with a negligible time jitter (0.2~ns) the 1PPmS signal
in NIM format.
The module output signals, seven for each spectrometer, are discriminated and 
also reshaped in NIM format.

Signals from RPC's and the 1PPmS are converted to differential ECL and sent to
a CAEN V767 VME TDC module (800~$\mu$s range, 0.8~ns Least Significant
Bit and integral non-linearity lower than 0.3~ns, multi-hits feature).
The TDC and the slave clock are acquired by means of a CAEN V1718 VME
bridge.
The acquisition system is triggered by a 4/14 majority of the RPC layers.

To obtain a ns--precision over the full range, the TDC was calibrated 
with a ppm precision.
For this purpose a thermally stabilized oscillator producing
a 5~kHz reference signal sent into a TDC channel was used during the whole data taking period.
This way, the stability of the TDC oscillator can be monitored event
by event, performing also the calibration with the required precision.
 

\section{Analysis}
\label{ana}

Each event is classified according to its topology~\cite{opcarac}: {\it CONTAINED} are events with the neutrino vertex in the target sections;
{\it SPECTRO} are events with the neutrino vertex in the spectrometer sections; {\it FRONTMUON} and {\it SIDEMUON} are events with the
neutrino vertex outside the detector and with a muon entering from the upstream and lateral sides of the detector, respectively.

Independent analyses were performed for events involving TT and RPC detectors.
Since $TOF_c$ (time of flight) is computed with respect to the origin of the OPERA reference frame which is located beneath the most upstream spectrometer magnet, the time measurements are corrected for the distance of the events along the beam line from this point, assuming time propagation at the speed of light.

Table~\ref{tab:events} summarizes the total number of events recorded, as well as the number of events used in the four analysis methods described in the next subsections.
Methods 1 and 2 were applied to TT data and Methods 3 and 4 to RPC data.

\begin{table}[hbt]
\begin{center}
\caption{Events used by the four analysis methods.}
\begin{tabular}{l c c c c c }
\hline
Event type & Events & Method 1 & Method 2 & Method 3 & Method 4 \\
\hline
{\it CONTAINED} & 18 & 17 & 13 & 13 & 11  \\
{\it FRONTMUON} & 37 & 29 & 27 & 27 & 24 \\
{\it SIDEMUON } & 17 & 3 & 3 & 9 & 8 \\
{\it SPECTRO } & 20 & 10 & 5 & 9 & 6 \\
Others & 12 & 0 & 0 & 0 & 0 \\ 
\hline 
{\it Total} & 104 & 59 & 48 & 58 & 49 \\ 
\hline
\end{tabular}
\label{tab:events}
\end{center}
\end{table}

For events where the muon crosses at least one spectrometer its momentum and charge are measured.
In 3 such events out of 58 CC events used in this analysis, the muon charge is positive.
One event is CONTAINED ($p_{\mu} \sim$12~GeV/c), the other ones are FRONTMUON events ($p_{\mu} \sim$46~GeV/c and $\sim79$~GeV/c). They exhibit a clear muon track crossing both spectrometers and thus all drift tube stations could be exploited for charge and momentum reconstruction.
The overall charge mis--identification probability for these events is below 1.5\%.
Considering also the charm background contribution, the number of expected fake positive muons from the whole BB sample is $< 0.7$ for the 3 observed events.
If the analysis is restricted only to contained events for which both charm contribution and charge mis--identification are lower, the expected number of fake positive muons is $< 0.07$ for 1 observed event.

\begin{figure}
  \centering
 \label{all_plots}
 \includegraphics[height=.50\textheight]{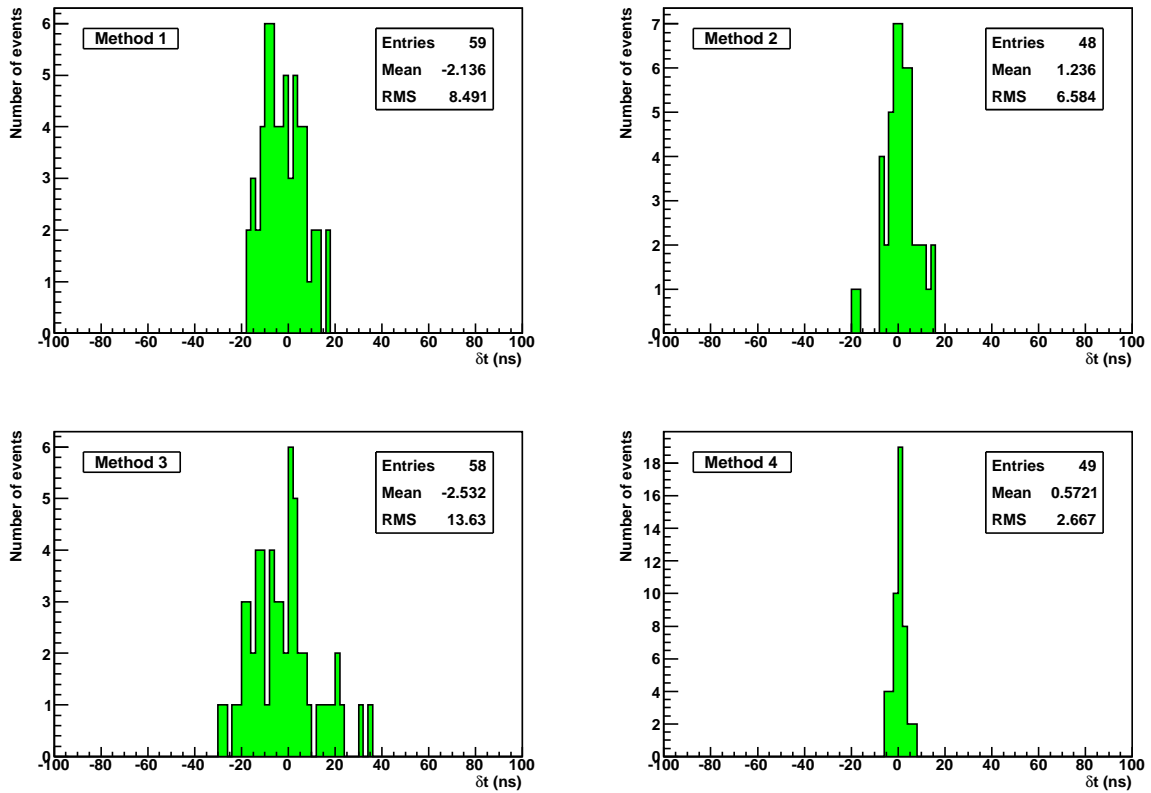}
 \caption{\small Neutrino time distribution of events selected for the TT analysis using the first hit (Method~1), for the TT analysis using all muon hits (Method~2), for events selected for the RPC analysis using all muon hits (Method~3) and for events selected for the RPC analysis using the TB (Method~4).}
\end{figure}

\subsection{Target Tracker data}

All 104 on--time recorded events were analyzed using the TT hit information, applying the same selection procedures as described in~\cite{v3}.

The selected data sample includes 63 events.
Four of them were rejected by the TDC cut to remove outliers (as reported in Section~\ref{ope}).

The neutrino interaction time and the corresponding time of flight $TOF_{\nu}$ were computed using two different methods.
The first method (Method~1) relies on the information of the earliest hit as described in~\cite{v3}.
The second method (Method~2) exploits the time information of all TT hits of the 3D muon track.
Using the latter method, 48 CC events were selected.
Method~2 results in a better time resolution (see last column of Tab.~\ref{tab:results}) due to the larger amount of information associated with the whole track.
However, the second method misses NC-like events (Neutral Current).
In Fig.~\ref{all_plots}, the upper left (right) plot shows $\delta t \equiv TOF_{c}-TOF_{\nu}$ for the first (second) method, respectively.
 The final result  is reported in Tab.~\ref{tab:results}.

\subsection{RPC data using standard DAQ}

For this method (Method~3), only neutrino interactions producing a clear muon track in the RPC's were used.
For tracks crossing both Super Modules an independent mean time was computed 
using the hits of the first and second Super Module separately.

The algorithm used to reconstruct the neutrino interaction time was the same as the one used for the TT analysis with Method~2.
The total number of events used with Method~3 is 58 (see Table~\ref{tab:events} for details).
The corresponding $\delta t$ distribution is shown in Fig.~\ref{all_plots} (Method~3) and the final result is given in Tab.~\ref{tab:results}.

\subsection{RPC Timing Board data}

With this new method (Method~4) the extra information provided by the special TDC (see Section~\ref{rpc_tb}) complements CNGS events acquired by the standard OPERA DAQ. 
The association of the additional data is performed on the basis of the UTC time readout by
the slave clock and of the VME TDC data.
In each event, accidental RPC hits inside the multi--hit 800~$\mu$s buffer of the VME TDC are rejected considering only those hits in a 128~ns window around the acquired trigger time.
The remaining hits acquired by the VME TDC are then corrected by the calibrated delays (for the position along the read-out strip, information from the DAQ is again used) and averaged.
The fit procedure is similar to the one used for Method~3.

The intrinsic time resolution computed using muons crossing both spectrometers is 1.1~ns.
This value contributes to the width of the distribution in the bottom right plot of Fig. 4 (Method~4), which is a convolution of the proton bunch spread and the uncertainty on the interaction point in external events.



\section{Results}
\label{res}

As discussed in the previous section, four different, correlated results were obtained for the neutrino velocity measurements using,
respectively:

\begin{enumerate}
	\item Standard DAQ, TT detectors and earliest hit timestamps,
	\item Standard DAQ and TT hit timestamps of the reconstructed muon track,
	\item Standard DAQ and RPC hit timestamps of the reconstructed muon track,
	\item Timing Board system and RPC detectors using muon hit timestamps.
\end{enumerate}

Methods 1 and 2 are statistically correlated since they use the same DAQ system and sub-detector.
They are also correlated with Method 3 using the same DAQ system.
Method 4 is almost uncorrelated with the other methods because of the different timing system used.
The residual correlation arises from the systematic error of the common part of the timing
chain.
Tab. \ref{sys_tab} reports the list of parameters and associated systematic errors of the different measurements.
The results obtained with each method are reported in Tab.~\ref{tab:results}, as well as the statistical and systematic errors.

In order to obtain a single combined result, Methods 2 and 4 were used
since they have the smallest statistical and systematic errors and - moreover - are almost completely uncorrelated
thus providing maximum information.
Values obtained for Method 1 were used for events when Method 2 could not be used because
of the absence of a muon track (20 events).
A standard combination procedure in presence of correlated
measurements was then used~\cite{lyons}.
The final result is: 

\begin{equation}
\delta t_\nu = ( 0.7 \pm 0.4\ (stat.) \pm 1.6\ (syst.-uncorr.) \pm 2.5\ (syst.-corr.) ) \ \textrm{ns}
\label{eq:nu1}
\end{equation}

Summing in quadrature the systematic errors, separately for $\nu$ and $\bar{\nu}$ contributions, leads to the results:
\begin{equation}
\delta t_\nu = ( 0.6 \pm 0.4\ (stat.) \pm 3.0\ (syst.) ) \ \textrm{ns}	
\label{eq:nu2}
\end{equation}
and:
\begin{equation}
\delta t_{\bar{\nu}} = ( 1.7 \pm 1.4\ (stat.) \pm 3.1\ (syst.) ) \ \textrm{ns}
\label{eq:antinu}
\end{equation}
for neutrinos and anti--neutrinos respectively (the slight difference in the systematic errors arises from the different contributions of each method to them).
When comparing $\delta t_\nu$ and $\delta t_{\bar{\nu}}$, systematic errors largely cancel out.
It was assumed that all the NC--like events result from $\nu$ interactions.
Since both results are compatible with zero, a limit on the deviation from the speed of light
was derived (90\%~C.L.):
\begin{equation}
-1.8 \times 10^{-6} < (v_{\nu}-c)/c < 2.3 \times 10^{-6}
\label{eq:limit}
\end{equation}
and:
\begin{equation}
-1.6 \times 10^{-6} < (v_{\bar{\nu}}-c)/c < 3.0 \times 10^{-6}
\label{eq:limit}
\end{equation}
for $\nu_{\mu}$ and $\bar{\nu}_{\mu}$, respectively.
It is pointed out that the statistical error in (5.3) was computed according to the RMS of the $\nu + \bar{\nu}$ distribution.
The above results are in agreement with those obtained by the other LNGS experiments participating to this CNGS BB run~\cite{others}.

\begin{table}
\begin{center}
\caption{Indivitual results for $\nu + \bar{\nu}$ for each of the four analysis methods reported in the text. 
In each row, the number of events, $\delta t \equiv TOF_c - TOF_\nu$, the statistical and systematic errors
and the time resolution on single events are reported (not including the 1.8~ns RMS of the proton bunches). All time values are in ns.}
\begin{tabular}{cccccc}
	\hline
	Method & Events & $\delta t$ & Stat. & Syst. & Resol. \\
	\hline
	     1 &   59   &   -2.1     &  1.1  &  4.4  &  9.8   \\
	     2 &   48   &    1.2     &  1.0  &  3.3  &  6.5   \\
	     3 &   58   &   -2.5     &  1.8  &  5.3  &  9.5   \\
	     4 &   49   &   0.6     &  0.4  &  3.6  &  1.1   \\
	\hline
\end{tabular}
\label{tab:results}
\end{center}
\end{table}

\section{Conclusions}
\label{con}

In May 2012 a two--week dedicated CNGS proton beam was provided to perform a measurement of the neutrino velocity.
The OPERA experiment after improving its timing system, has confirmed the result reported in~\cite{v3}, showing no significant deviation of the muon neutrino velocity from the speed of light.
The present result is $\delta t = 0.6\pm 0.4\ (stat.) \pm 3.0 \ (syst.)$~ns, giving the limit $-1.8 \times 10^{-6} < (v_{\nu}-c)/c < 2.3 \times 10^{-6}$ at 90\%~C.L.
During the same period, 3 muon anti--neutrino interactions were also recorded.
The corresponding $\delta t $ value is $1.7\pm 1.4\ (stat.) \pm 3.1 \ (syst.)$~ns compatible with that obtained with neutrino events.

\section{Acknowledgements}
\label{ack}

We thank CERN for the successful operation of the accelerator complex and the CNGS facility, and for the prompt setting up of the bunched proton beam. We are indebted to INFN for the continuous support given to the experiment during the construction, installation and commissioning phases through its LNGS laboratory. Funding from our national agencies is warmly acknowledged: Fonds de la Recherche Scientifique - FNRS and Institut Interuniversitaire des Sciences Nucl\'eaires for Belgium; MoSES for Croatia; CNRS and IN2P3 for France; BMBF for Germany; INFN for Italy; JSPS (Japan Society for the Promotion of Science), MEXT (Ministry of Education, Culture, Sports, Science and Technology), QFPU (Global COE program of Nagoya University, ``Quest for Fundamental Principles in the Universe'' supported by JSPS and MEXT) and Promotion and Mutual Aid Corporation for Private Schools of Japan for Japan; the Swiss National Science Foundation (SNF), the University of Bern and ETH Zurich for Switzerland; the Russian Foundation for Basic Research (grant 09-02-00300 a), the Programs of the Presidium of the Russian Academy of Sciences ``Neutrino Physics'' and ``Experimental and theoretical researches of fundamental interactions connected with work on the accelerator of CERN'', the Programs of support of leading schools (grant 3517.2010.2), and the Ministry of Education and Science~ of~ the Russian Federation for Russia; the Korea Research Foundation Grant (KRF-2008-313-C00201) for Korea; and TUBITAK The Scientific and Technological Research Council of Turkey, for Turkey. We are also indebted to INFN for providing fellowships and grants to non-Italian researchers. We thank the IN2P3 Computing Centre (CC-IN2P3) for providing computing resources for the analysis and hosting the central database for the OPERA experiment.
For the measurement of the neutrino time of flight with the timing boards we thank the LNF ROG group, the electronics workshops of LNF and LNGS for the finalization of the VME clock slave and the LNGS computing center for the support in the calibration of the 1PPmS signal fiber delay.


\end{document}